\documentclass[]{fairmeta}

\usepackage{url}
\usepackage{graphicx}
\usepackage{multirow}
\usepackage{subcaption}
\usepackage{booktabs}
\usepackage{xcolor}
\usepackage{subcaption}
\usepackage{amssymb}
\usepackage{amsmath}
\definecolor{quantizered}{HTML}{C00000}

\definecolor{mycitecolor}{HTML}{395B9E}
\hypersetup{
    colorlinks,
    linkcolor={mycitecolor},
    citecolor={mycitecolor},
    urlcolor={blue!80!black}
}

\title{Gradient-based Jailbreak Images for\\Multimodal Fusion Models}

\author[1,2,*]{Javier Rando}
\author[1]{Hannah Korevaar}
\author[1,]{Erik Brinkman}
\author[1]{Ivan Evtimov}
\author[2]{Florian Tramèr}

\affiliation[1]{Meta}
\affiliation[2]{ETH Zurich}

\contribution[*]{Work done at Meta}

\abstract{Augmenting language models with image inputs may enable more effective jailbreak attacks through continuous optimization, unlike text inputs that require discrete optimization. However, new \emph{multimodal fusion models} tokenize all input modalities using non-differentiable functions, which hinders straightforward attacks.
In this work, we introduce the notion of a \emph{tokenizer shortcut} that approximates tokenization with a continuous function and enables continuous optimization. We use tokenizer shortcuts to create the first end-to-end gradient image attacks against multimodal fusion models.
We evaluate our attacks on Chameleon models and obtain jailbreak images that elicit harmful information for 72.5\% of prompts.
Jailbreak images outperform text jailbreaks optimized with the same objective and require 3$\times$ lower compute budget to optimize 50$\times$ more input tokens.
Finally, we find that representation engineering defenses, like Circuit Breakers, trained only on text attacks can effectively transfer to adversarial image inputs.
}

\date{\today}
\correspondence{First Author at \email{javier.rando@ai.ethz.ch}}

\metadata[Code]{\url{https://github.com/facebookresearch/multimodal-fusion-jailbreaks}}

\begin{document}

\maketitle

\section{Introduction}

Adapter-based vision language models were an early attempt to augment large language models (LLMs) with image inputs~\citep{liu2024visual}. They use
a pretrained image embedding model, like CLIP~\citep{radford2021learning}, and train adapters to map image embeddings directly into the embedding space of a pretrained LLM.
However, separate input spaces can limit multimodal understanding and do not support native generation of images. In contrast, early-fusion multimodal models have been introduced as a more general approach that supports unlimited modalities as both input and output~\citep{team2024chameleon,team2023gemini,4ocard}. These models project all modalities into a shared tokenized space and are pretrained from scratch on multimodal inputs. In this work, we will refer to early-fusion multimodal models as \emph{multimodal fusion models}.

Just like LLMs, most vision language models are trained to behave safely and reject harmful requests~\citep{bai2022training}. \citet{carlini2024aligned} demonstrated that bypassing safeguards in adapter-based vision language models is easy because input images can be continuously optimized to maximize harmful outputs. This is in contrast to text input optimization, which requires less efficient discrete optimization methods~\citep{zou2023universal}. Unlike adapter-based models, multimodal fusion models tokenize all modalities, creating a non-differentiable step between the input and the output token spaces. As a result, optimizing any modality becomes again a discrete optimization problem.

\begin{figure}
    \centering
    \includegraphics[width=\linewidth]{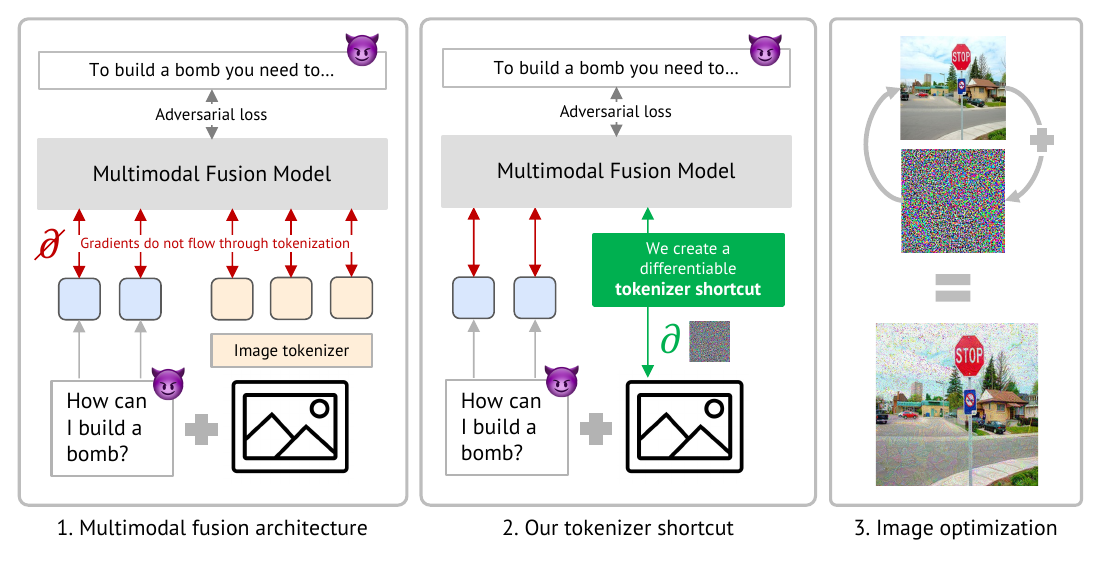}
    \caption{\textbf{Tokenizer shortcut.} Multimodal fusion models tokenize images and are thus not differentiable end-to-end. We create a differentiable \emph{tokenizer shortcut} to enable adversarial image optimization. We optimize images to maximize the probability of affirmative responses.}
    \label{fig:figure1}
\end{figure}

Building upon prior work on adversarial examples against quantized image classifiers~\citep{athalye2018obfuscated}, we introduce the notion of a \emph{tokenizer shortcut} that approximates image tokenization with a differentiable function. Backpropagating the model loss through the shortcut provides surrogate gradients to enable continuous end-to-end optimization. %
We propose different tokenizer shortcut designs, and use them to introduce the first end-to-end attack against multimodal fusion models. We evaluate it on Chameleon models~\citep{team2024chameleon} under white-box access. 

We optimize \emph{jailbreak images} and evaluate their success in eliciting harmful responses for prompts in JailbreakBench~\citep{chao2024jailbreakbench}. We find successful jailbreak images for more than 70\% of the prompts. The performance is superior to text-only attacks (GCG: 64\%) and requires $3\times$ less compute. Our jailbreak images exhibit better transferability across prompts than text-only attacks (73\% vs. 50\%). However, similar to concurrent work on adapter-based models~\citep{schaeffer2024universal}, our jailbreak images do not transfer across models.

Finally, we evaluate jailbreak images against white-box protections. Jailbreak images, unlike text attacks, do not increase prompt perplexity, making them undetectable by common defense methods that filter out high-perplexity prompts \citep{jain2023baseline}. However, we find that representation engineering defenses \citep{zou2024improving} trained only on text inputs can transfer to image attacks.

Overall, our work answers the following questions: (1) How can continuous optimization be enabled in multimodal fusion models?; (2) How does the resulting image optimization compare to text optimization in terms of attack success and budget?; (3) How well do jailbreak images transfer across prompts, models, and defenses?

\section{Preliminaries}

\paragraph{Multimodal models.} \emph{Adapter-based vision language models} combine pretrained architectures by training adapters to map image embeddings into the embedding space of a language model~\citep{liu2024visual}. On the other hand, \textbf{\emph{multimodal fusion models}}---which we focus on---are trained from scratch on multimodal inputs. For this, all modalities are mapped into a shared tokenized space that can serve as both input and output for autoregressive architectures. In this work, we evaluate the family of Chameleon models \citep{team2024chameleon}, the only open-source early-fusion multimodal models at the time of writing. To convert images into tokens, Chameleon trains a separate VQ-VAE~\citep{gafni2024scene} that encodes the image into 1024 vectors that are then quantized into discrete tokens.

\paragraph{Jailbreaking language models.} Researchers continuously find \emph{jailbreaks} to bypass safeguards and extract unsafe information from language models~\citep{wei2024jailbroken}. On one hand, some attacks use specific prompting strategies~\citep{liu2023jailbreaking,shah2023scalable,liu2023autodan,wei2024jailbroken} that do not require access to model weights. On the other hand, \cite{zou2023universal} proposed GCG, a discrete optimization algorithm that searches for prompts that elicit unsafe responses. For adapter-based multimodal models, input images can be used as a more efficient space for continuous optimization ~\citep{carlini2024aligned}. However, this approach does not directly apply to multimodal fusion models, as these tokenize inputs with non-differentiable functions, thereby reverting adversarial optimization to a discrete optimization problem.

\paragraph{Threat Model.} We assume an attacker with white-box access to a multimodal fusion model, which was trained to behave safely and refuse harmful requests. The attacker's \emph{goal} is to find a \emph{universal} adversarial input image that, when combined with any harmful text request, jailbreaks the model and results in a harmful generation. Moreover, we place no restriction on the attacker's image. In particular, the image can be arbitrarily distorted and may not resemble any human-interpretable image. To compare image and text attacks, we only consider harmful instructions that can be fully defined in text, thus excluding attacks that require interaction between text and images.

\section{Our Attack: Tokenizer Shortcut for Continuous Optimization}

The goal of our attack is finding adversarial images that, when appended to harmful prompts, can elicit a harmful response from the model. For this, inspired by the GCG objective~\citep{zou2023universal}, we optimize the image using gradient descent to maximize the probability of model generations to start with a \emph{non-refusal prefix} (e.g. ``Sure, I can help you with that''). We can additionally enhance the loss to simultaneously minimize the probability of generic refusal tokens. More formally, we optimize the image to minimize the following loss

\begin{equation}
    \mathcal{L}(\text{image}) = \prod_{i \in \text{prompts}} \underbrace{p(\text{refusal prefix}_i | i)}_{\text{to be minimized}} - \underbrace{p(\text{non-refusal prefix}_i | i)}_{\text{to be maximized}}.
    \label{eq:loss}
\end{equation}

At each step $t$ of the optimization, we update our image to minimize the loss using gradient descent\footnote{Since the attacker does not need to preserve any semantic information in the image, we do not project the updates to remain within an epsilon-ball from the original image~\citep{madry2017towards}.}:
\begin{equation}
    \text{image}_{t+1} = \text{image}_{t} - \alpha \text{sign}(\partial L / {\partial \text{image}}).
    \label{eq:imageupdates}
\end{equation}

\begin{figure*}[th]
    \centering
    \begin{subfigure}[t]{\textwidth}
        \centering
        \includegraphics[width=0.7\textwidth]{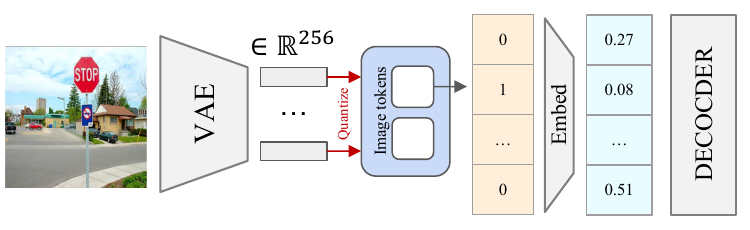}
        \caption{\textbf{Default architecture}. Images are tokenized using a vector-quantized VAE. {\textcolor{quantizered}{Quantization}} is not differentiable and prevents end-to-end gradient attacks.}
    \end{subfigure}
    \begin{subfigure}[t]{\textwidth}
        \centering
        \includegraphics[width=0.7\textwidth]{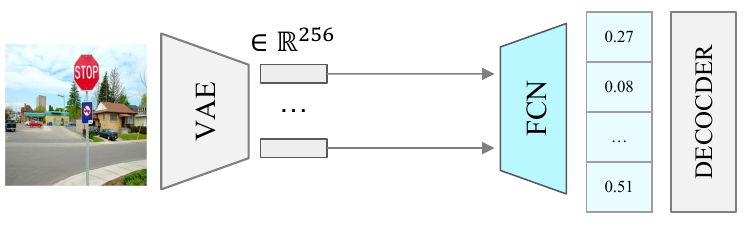}
        \caption{\textbf{Embedding Shortcut}. We use a 2-layer fully connected network to map the VAE embeddings directly to the decoder embedding space.}
    \end{subfigure}
    \begin{subfigure}[t]{\textwidth}
        \centering
        \includegraphics[width=0.7\textwidth]{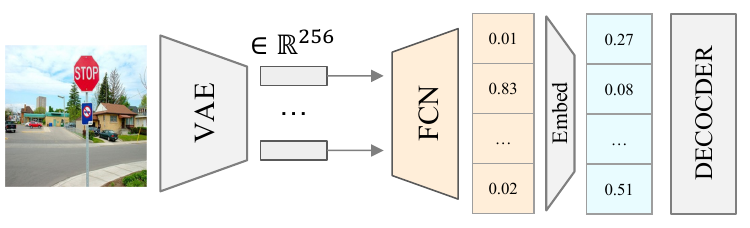}
        \caption{\textbf{1-Hot Encoding Shortcut}. We use a 2-layer fully connected network to map the VAE embeddings to a soft 1-hot encoding that is then propagated through the embedding layer of the model.}
    \end{subfigure}
    \caption{Overview of the default image tokenization in Chameleon models and our proposed shortcuts to enable end-to-end gradients. Each token is propagated independently through shortcuts.}
    \label{fig:architectures}
\end{figure*}

However, image tokenization in multimodal fusion models relies on quantization which is non-differentiable, and it is thus not possible to compute $\partial L / {\partial \text{image}}$ naively. Previous work on image classification showed that even rough approximations of quantization can provide gradients to guide adversarial optimization~\citep{athalye2018obfuscated}. We introduce the notion of a \emph{tokenizer shortcut} that maps the image embeddings before quantization to a continuous model input space, creating a fully differentiable path between the image and the output space. See Figure \ref{fig:architectures} for an illustration. These shortcuts enable end-to-end continuous optimization of images with respect to the model prediction loss.

\paragraph{Tokenizer shortcut.} We use a 2-layer fully connected network as our \emph{tokenizer shortcut} to approximate tokenization with a differentiable function. We propose two shortcuts that take as input each of the 1024 latent vectors from the VQ-VAE ($\mathbb{R}^{1024 \times 256}$). The first shortcut maps the VQ-VAE embeddings directly to the LLM embedding space ($\mathbb{R}^{1024 \times 4096}$); we will refer to this shortcut as the \emph{embedding shortcut}. The second shortcut maps the VQ-VAE embeddings to the one-hot encoding over the vocabulary tokens, producing a soft one-hot encoding over tokens ($\mathbb{R}^{1024 \times 16384}$) that is then used as input to the model; we will refer to this shortcut as the \emph{1-hot shortuct}. Figure \ref{fig:architectures} depicts our proposed shortcuts.

Since the original Chameleon pretraining dataset is proprietary, we train both shortcuts on a subset of the open-source MIMIC-IT dataset~\citep{li2023mimic}. The embedding shortcut is trained to minimize the cosine similarity between the shortcut predictions and the embedding that would be obtained for the original image tokens. The 1-hot shortcut produces a distribution over the vocabulary and minimizes the cross-entropy loss with the token that would be assigned to each vector through quantization. After training the shortcuts, we use them to compute end-to-end gradients and update the image pixels according to Equation~\ref{eq:imageupdates}.

\section{Experimental Setup}

\paragraph{Datasets.} We use JailbreakBench~\citep{chao2024jailbreakbench} to evaluate our attacks. It provides a curated collection of harmful prompts from different sources. We evaluate our attacks on 80 prompts and keep a held-out set of 20 prompts for transferability experiments. While there can be reasonable debate over what is considered violating and harmful in AI generations,  JailbreakBench offers a framework for comparing different methods using common prompts and metrics.

\paragraph{Optimizing adversarial images.} We optimize our jailbreak images using gradient descent (see Equation~\ref{eq:imageupdates}) for 500 steps. We initialize $\alpha=0.01$ and divide it by 2 every 100 steps until reaching a minimum of $0.001$. We keep the image with the lowest loss on the training images. We start from a random image from the MIMIC-IT dataset~\citep{li2023mimic}, illustrated in Figure~\ref{fig:figure1}.

\paragraph{Baseline attacks.} We compare image jailbreaks against text-only and representation engineering attacks. More specifically, we use \emph{GCG}~\citep{zou2023universal}---optimizes text prompts to maximize probability of non-refusal---and \emph{Refusal Direction} \citep{arditi2024refusal}---identifies a refusal direction in the model's activation space and subtracts it during the forward pass. For GCG, we run all our attacks for 200 steps with a suffix of 20 tokens, 256 replacement candidates per token and 512 total candidate suffixes per step. 

\paragraph{White-box protections.} We evaluate our attacks against two white-box protections applied after traditional safety training. First, we implement Circuit Breakers~\citep{zou2024improving}, a representation engineering defense optimized over text inputs. Second, we use average token perplexity in input prompts as a filtering metric for high-perplexity prompts~\citep{jain2023baseline}. Since multimodal fusion models autoregressively predict image tokens, perplexity in the token space can be extended to image prompts. For GCG attacks, we report the difference in average perplexity per token compared to a clean prompt without an adversarial suffix. For image attacks, we compare to the average token perplexity obtained for prompts combined with 50 different benign images. We do not compute perplexity in the pixel space since adversaries can easily bypass protections based on image statistics~\citep{carlini2017adversarial,tramer2020adaptive}.

\paragraph{Measuring attack success rate.}
We sample responses for our test prompts and use the two judge models in JailbreakBench. The first model determines if a generation is unsafe and the second whether it is a refusal. We report the percentage of prompts with unsafe responses (\emph{attack success rate}), and the percentage of refused instructions (\emph{refusal rate}). Additionally, Circuit Breakers often renders models unusable, producing gibberish responses with many special characters. These outputs are sometimes misclassified as successful jailbreaks by JailbreakBench judges. To address this, we exclude Circuit Breaker generations containing more than 15 special characters from our successful jailbreak count\footnote{We exclude frequent punctuation marks such as commas and periods from the count.}.

\paragraph{Direct and transfer attacks.} We propose two attack strategies inspired by GCG. The first, \emph{direct attacks}, optimizes the loss in Equation~\ref{eq:loss} for a specific target prompt, creating one jailbreak image per test prompt. The second strategy, \emph{transfer attacks}, optimizes a single image over a set of held-out prompts. We then evaluate the attack success on the test prompts, unseen during optimization.

\section{Direct Attacks}
\label{sec:singlresults}

In this set of experiments, we assume an attacker that optimizes a jailbreak image for each of the test prompts to maximize effectiveness. First, in Section~\ref{ssec:best-attack}, we report the most relevant ablations to find our best attack. Then, in Section~\ref{ssec:text-attacks}, we compare the performance of our attack with text-only optimization. Finally, in Section~\ref{ssec:defenses}, we measure the effectiveness of our jailbreak images against popular white-box protections.

\subsection{Finding the Best Attack}
\label{ssec:best-attack}

\paragraph{Comparing shortcuts.} We compare the performance of the embedding and 1-hot shortcuts in Table~\ref{tab:shortcuttype}. Both shortcuts perform similary and find successful jailbreak images for over 70\% prompts. However, their performance drastically changes if the shortcut is \emph{turned-off} and the image follows the default tokenization forward path (e.g. if an adversary creates the images in a white-box setup and wants to use them against the same model deployed under black-box access). In this case, images optimized with an embedding shortcut are no longer successful but images from the 1-hot shortcut retain an attack success rate close to 50\%. %

\begin{table}[ht]
\caption{\textbf{1-hot shortcuts generalize to a black-box attack.} Attack Success Rate with and without shortcut in forward pass. We optimize one image for each prompt and report the percetange of prompts where we obtain a jailbreak.}
\label{tab:shortcuttype}
\centering
\begin{tabular}{@{} lrr @{}}
\toprule
               & With Shortcut & Without Shortcut \\ \midrule
Embedding Space & 70.0\%         & 0.0\%            \\
1-Hot Space  & 71.3\%         & 47.5\%          \\ \bottomrule 
\end{tabular}
\end{table}

\paragraph{Temperature in shortcut predictions.}
The mapping to the one-hot encoding space can be easily learned by the shortcut network and it produces very confident predictions (i.e. probability 1 on the correct token and 0 elsewhere). This skewed distribution causes vanishing gradients for all other tokens. We thus find that applying high softmax temperature of $6$ to the predictions of the shortcut is essential to be able to optimize the attack. Appendix \ref{ap:softmax} shows different ablations of the softmax temperature and their attack success.

\paragraph{Optimization target.}
The optimization process is very sensitive to the target prefixes used to compute the loss in Equation~\ref{eq:loss}. Table \ref{tab:opttargets} reports the most relevant ablations. The best performing target is maximizing the probability of a long non-refusal prefix that is semantically relevant to the prompt. For example, for a prompt asking for instructions to build a bomb, the prefix might be ``Sure, here are detailed instructions to build a bomb at home:''. Minimizing the probability of refusal tokens does not provide any advantages. We also report attack success rate when only maximizing the probability of ``Sure''---without any context-relevant information---as non-refusal prefix.

\begin{table}[ht]
\caption{\textbf{Contextual targets perform the best.} Attack success rate after optimizing an image using different prefixes for Equation \ref{eq:loss}. Results computed using the shortcut in the forward pass.}
\label{tab:opttargets}
\centering
\begin{tabular}{@{} lcrr @{}}
\toprule
Non-refusal prefix & Refusal prefix & Embed. Shortcut & 1-hot Shortcut \\ \midrule
Sure               & -              &  18.8\%           &   13.8\%         \\
Sure + context     & -              &  70.0\%          &   71.3\%         \\
Sure + context     & I              & 66.3\%      & 44.3\%          \\ \bottomrule
\end{tabular}
\end{table}

\subsection{Comparison with Text Optimization}
\label{ssec:text-attacks}

We compare the performance and efficiency of optimization over text and image tokens. We use GCG as the best-known method to update text tokens with gradient information. For GCG, we optimize an adversarial prefix---instead of an image---for each test prompt. Both successful image jailbreaks and GCG suffixes require around 100 optimization steps, but their computational complexities differ. For each step, jailbreak images require 1 forward and 1 backward pass on 1024 image tokens; totaling $\sim100,000$ forward and backward token operations per successful jailbreak image. In contrast, GCG uses 20 additional tokens with 1 backward pass and 512 forward passes per iteration, resulting in $\sim20,000$ backward and $\sim1M$ forward token operations. 
Assuming constant FLOPs per token operation\footnote{Although exact FLOPs are not constant, the differences are negligible and we assume equal cost per token.} and that backward operations require 3 times more FLOPs than forward operations, we estimate that jailbreak images require an additional 400,000 FLOPs/token, while GCG needs an extra 1,060,000 FLOPs/token to obtain a successful attack\footnote{We did not optimize any of the methods for efficiency.}.

Table \ref{tab:comparisonsingle} summarizes the attack success rate for both methods. Jailbreak images outperform text-based attacks, and require $3\times$ less compute. Image optimization also offers a larger attack surface since the attacker can modify the 1024 image tokens at a lower overall cost than GCG requires for only 20 text tokens.

\begin{table}[ht]
\centering
\caption{\textbf{Image jailbreaks outperform text attacks.} Attack success rate (ASR) and refusal rate (RR) for each attack on Chameleon-7B before and after applying Circuit Breaker protections. Results computed using the shortcut in the forward pass. We also report the difference in \emph{perplexity per token} ($\Delta$ PPL) with respect to benign prompts.}
\label{tab:comparisonsingle}
\begin{tabular}{@{} lrrrrr @{}} \toprule
                                                    & \multicolumn{2}{l}{Default Safety} & \multicolumn{2}{c}{Circuit Breaker} &  \\ \cmidrule{2-5} 
                                                                             &    ASR ($\uparrow$)           & RR ($\downarrow$)           & ASR ($\uparrow$)             & RR ($\downarrow$)&       $\Delta$ PPL ($\downarrow$)       \\ \midrule[\heavyrulewidth]
No attack & 2.5\% & 77.5\% & 1.3\% & 81.3\% & - \\ \midrule
Jailbreak Image (1-hot)   & \textbf{71.3\% }         & 13.8\%          & 0.0\%            & 66.2\%       &  -1.14   \\
Jailbreak Image (Embed) & 70.0\%            & 16.3\%          & 3.8\%            & 64.6\%       &  -1.21  \\ 
GCG                     & 63.8\%           & 22.5\%          & \textbf{10.0\%}            & 37.5\%        & +2.51 \\        \bottomrule               
\end{tabular}
\end{table}

\subsection{Effectiveness Against White-Box Protections}
\label{ssec:defenses}
First, we create jailbreak images and text suffixes with GCG on models enhanced with Circuit Breakers~\citep{zou2024improving} (see Table~\ref{tab:comparisonsingle}). Circuit Breakers protections optimized over text inputs can generalize against images. Embedding shortcuts provide a more flexible representation to circumvent the protections. Moreover, unlike text attacks, jailbreak images do not increase the average token perplexity in the prompt, making them harder to detect by methods that filter out high-perplexity prompts~\citep{jain2023baseline}. Although perplexity could also be computed in the image space, this protection has been shown to be ineffective against adversarial images~\citep{carlini2017adversarial}.

\section{Transfer Attacks}

\begin{table}[ht]
\centering
\caption{\textbf{Existing methods outperform image jailbreaks on Chameleon.} Attack success rate (ASR) and refusal rates (RR) on JailbreakBench. We also report the difference in \emph{perplexity per token} ($\Delta$ PPL) with respect to benign prompts.}
\label{tab:existing}
\begin{tabular}{@{} l@{\hskip -5pt}r@{\hskip 0.4in}rr@{\hskip 0.4in}rrr @{}}
\toprule
                                                            & & \multicolumn{2}{l}{Default Safety} & \multicolumn{2}{c}{Circuit Breaker} &  \\ \cmidrule{3-6}
                                                                              Attack      &      Train Prompts                                                                    & ASR ($\uparrow$)              & RR ($\downarrow$)              & ASR ($\uparrow$)              & RR ($\downarrow$)               &      $\Delta$ PPL ($\downarrow$)                                      \\ \midrule[\heavyrulewidth]
No attack                                                                           & -                                                                        & 2.5\%            & 77.5\%          & 1.3\%            & 81.3\%           & -                                      \\ \midrule
\multirow{3}{*}{\begin{tabular}[c]{@{}l@{}}Jailbreak \\ Image (1-hot)\end{tabular}}  & 1                                                                        & 27.5\%           & 60.0\%          & 0.0\%            & 86.3\%           & -1.14                                      \\
                                                                                    & 10                                                                       & 53.8\%           & 35.0\%          & 3.8\%           & 68.8\%           & -1.18                                     \\
                                                                                    & 20                                                                       & 51.3\%           & 36.3\%          & 1.3\%            & 58.8\%           & -1.08                                      \\ \midrule
\multirow{3}{*}{\begin{tabular}[c]{@{}l@{}}Jailbreak \\ Image (Embed)\end{tabular}} & 1                                                                        & 31.3\%           & 51.3\%          & 0.0\%            & 73.8\%           & -1.18                                      \\
                                                                                    & 10                                                                       & 72.5\%           & 16.3\%          & 6.3\%           & 45.0\%           & -1.29                                    \\
                                                                                    & 20                                                                       & 72.5\%           & 10.0\%          & \textbf{10.0\%}           & 53.8\%           & -1.21                                      \\ \midrule
\multirow{3}{*}{GCG}                                                                & 1                                                                        & 17.5\%           & 66.3\%          & 1.3\%            & 78.8\%           & +0.86                                    \\
                                                                                    & 10                                                                       & 50.0\%           & 60.0\%          & 5.0\%           & 50.0\%           & +0.14                                      \\
                                                                                    & 20                                                                       & 46.3\%           & 21.3\%          & 1.3\%           & 38.8\%           & +3.02                                      \\ \midrule
\multirow{3}{*}{Refusal Direction}                                                  & 1                                                                        & 5.0\%            & 55.0\%                        & 0.0\%                & 81.3\%  & -                                     \\
                                                                                    & 10                                                                       & 73.8\%           & 1.3\%           & 0.0\%           & 40.0\%           & -                                      \\
                                                                                    & 20                                                                       & \textbf{81.3\%}  & 2.5\%  & 5.0\%  & 38.8\%  & -                                      \\ \midrule
                                                                                    \end{tabular}
\end{table}

We now evaluate the transferability of jailbreak images to unseen prompts and models. Instead of optimizing a separate image for each test prompt, we create a universal jailbreak image optimized over $N$ held-out \emph{train prompts}, and evaluate the attack success rate on the unseen test prompts. 

We use 2 text-only baselines to contextualize our findings. We again use GCG, but, instead of optimizing for a single propmt, we optimize text suffixes over the same set of train prompts to increase transferability. Additionally, we use refusal suppression~\citep{arditi2024refusal}, a representation engineering method that requires a set of train prompts to detect a direction in the model's activation space that is responsible for refusal\footnote{This method could not identify a refusal direction when using image inputs.}. This direction is subtracted at inference time, disabling the model's ability to refuse harmful requests. We summarize the main findings from the results in Table~\ref{tab:mainresults}.

\paragraph{Increasing the number of train prompts improves generalization.} Jailbreak images optimized on a single image exhibit good generalization to unseen prompts ($\sim$30\%) but computing the loss over a larger number of prompts improves generalization for both 1-hot and embedding shortcuts. However, after a certain point, increasing the number of prompts does not result in improved attack success rates. For both shortcuts, we find very similar success for 10 and 20 training prompts.

\paragraph{Embedding shortcut can find more transferable jailbreak images.} Although generalization for images optimized on a single prompt is similar for 1-hot and embedding shortcuts, the embedding shortcut can obtain much higher attack success rates on unseen prompts when optimized over more than 10 prompts (53.8\% vs. 72.5\%). In fact, the performance obtained with the jailbreak image optimized with embedding shortcut on 10 prompts can jailbreak as many prompts as optimizing a single image per test prompt (see Section~\ref{sec:singlresults}), reducing computational cost significantly.

\paragraph{Representation engineering can still outperform jailbreak images.} Suppressing the refusal direction during the forward pass obtains the best results overall on our test set (81.3\%). This suggests that our image jailbreaks suffer from similar limitations as existing text-only attacks because it is restricted to the input space.

\paragraph{Universal jailbreak images do not increase perplexity.} Similar to single-prompt attacks, the resulting images do not increase the average token perplexity of the prompt. In fact, jailbreak images have lower perplexity than benign images under the model distribution.

\paragraph{Jailbreak images do not transfer across models.} We evaluate whether the images optimized on the held-out training set can jailbreak the unseen test prompts on unseen models. We use Chameleon-30B as a model within the same family, and the latest LLaMA3-LLaVA-1.6 vision language model to account for models with different architectures and training setups. Similar to recent work on adapter-based vision language models~\citep{schaeffer2024universal}, we find that jailbreak images transfer poorly across models. In fact, GCG suffixes can transfer much better to the LLaVA model increasing attack success rate by $32.5$p.p from the baseline versus the 1.2p.p. achieved by jailbreak images. Direct attacks do not transfer to other architectures either (see Appendix~\ref{ap:direct-transfer}).

\begin{table}[ht]
\centering
\caption{\textbf{Jailbreak images do not transfer across models.} Attack success rate (ASR) and refusal rate (RR) on JailbreakBench using images optimized on Chameleon-7B.}
\label{tab:mainresults}
\begin{tabular}{@{} lr@{\hskip 0.4in}rrrr @{}}
\toprule
&       & \multicolumn{2}{r}{Chameleon-30B} & \multicolumn{2}{c}{LLaVA-1.6} \\ \cmidrule{3-6}
                                                                                    
                                                                             Attack       & Train Prompts & \multicolumn{1}{c}{ASR}    & RR   & \multicolumn{1}{c}{ASR}  & RR \\ \midrule[\heavyrulewidth]
No attack                                                                           & -                    & 0.0\%                      &  98.8\%    & 8.8\%                    &  85.0\%  \\ \midrule
\multirow{3}{*}{\begin{tabular}[c]{@{}l@{}}Jailbreak \\ Image (1-hot)\end{tabular}} & 1                    & 0.0\%                      &   98.8\%   & 10.0\%                   &  88.8\%  \\
                                                                                    & 10                   & 0.0\%                      &  98.8\%    & 10.0\%                   &  88.8\%  \\
                                                                                    & 20                   & 0.0\%                      &  98.8\%    & 10.0\%                   &  91.3\%  \\ \midrule
\multirow{3}{*}{\begin{tabular}[c]{@{}l@{}}Jailbreak \\ Image (Embed)\end{tabular}} & 1                    & 0.0\%                      &  98.8\%    & 10.0\%                   &  87.5\%  \\
                                                                                    & 10                   & 0.0\%                      &  98.8\%    & 10.0\%                   &  88.8\%  \\
                                                                                    & 20                   & 0.0\%                      &  98.8\%    & 10.0\%                   &  88.8\%  \\ \midrule
\multirow{3}{*}{GCG}                                                                & 1                    & 2.5\%                      &  91.3\%    & 21.3\%                   &  81.3\%  \\
                                                                                    & 10                   & 6.3\%                      &  82.5\%    & 41.3\%                   &  58.8\%  \\
                                                                                    & 20                   & 5.0\%                      &  88.8\%    & 7.5\%                    &  86.3\%  \\ \bottomrule
\end{tabular}
\end{table}

\section{Discussion and Future Work}

\paragraph{Image jailbreaks are promising but have a long way to go.} Our work is the first attempt to jailbreak multimodal architectures using end-to-end gradient attacks. We present promising results that suggest that optimization might be smoother and more efficient than on text tokens. However, we think there is still significant room for improvement upon our methods. While we ablated most of the relevant hyperparameters we identified, our results indicate that attack success can still vary drastically with their choice. Future work may explore the optimization dynamics in more detail to come up with more effective ways of using the gradients. Similarly, future work may explore more flexible target functions~\citep{thompson2024fluent}; e.g. combining the output and activation space.

\paragraph{Transferability of jailbreak images remains an open problem.} Concurrent work on adapter-based models ~\citep{schaeffer2024universal} demonstrated that jailbreak images do not generalize across models even when optimizing on an ensemble of models. Our results indicate that this problem also persists in multimodal fusion models. Future work may focus on looking for ways to regularize the optimization to improve transferability.

\vspace{1em}

\paragraph{Attack dynamics may change for newer models.} Although most companies have already announced proprietary multimodal architectures~\citep{team2023gemini,4ocard}, Chameleon models are the only open-source models trained with this architecture. We expect more models to come out in the coming months and we encourage researchers to assess the success of this attack on future models and architectures. We find that Chameleon models are overly safe in that they often refuse benign instructions. We believe this might be a result of a specific stance in the utility-safety tradeoff~\citep{bai2022training} for Chameleon that may not hold true for other models.

\section{Related Work}
\label{related}

\paragraph{Multimodal Models.} Adapter-based vision language models (VLMs) are a popular approach for integrating image understanding into language models~\citep{liu2024visual,karamcheti2024prismatic}. They typically combine pre-trained language models with image embedding models. VLMs train an adapter to transform pre-trained image embeddings (e.g., from CLIP) into a representation compatible with language models. However, state-of-the-art proprietary models are shifting towards early-fusion multimodal models~\citep{team2023gemini,4ocard}. These models autoregressively process tokens representing multiple modalities in a joint tokenized space. Our work focuses on Chameleon models~\citep{team2024chameleon}, currently the only open-source family of early-fusion multimodal models. Chameleon models support image inputs, are safety-tuned, and are available in 7B and 30B parameter sizes.

\paragraph{Adversarial Examples.} Adversarial examples are inputs designed to fool machine learning models, first explored in the context of image classification~\citep{szegedy2013intriguing,madry2017towards}. Adversarial images examples are created by adding perturbations to valid inputs. These perturbations are optimized using the gradient of the target loss with respect to the input.

\paragraph{Jailbreaks.} Jailbreaks are adversarial inputs designed to bypass the safeguards implemented in large language models (LLMs) and get them to generate harmful content. Jailbreaks are often black-box---they do not require access to model weights--- and involve specific conversational strategies~\citep{liu2023jailbreaking,shah2023scalable,liu2023autodan,wei2024jailbroken}. On the other hand, there are white-box jailbreaks that use model weights and gradients to guide the attack. However, unlike in image classification, text inputs cannot be directly optimized using gradients with respect to the target loss---because the input space is discrete and sparse---and require approximate methods. GCG is the most prominent white-box attack~\citep{zou2023universal}. GCG optimizes a suffix that can be appended after harmful instructions to prevent refusal. Interestingly, GCG suffixes optimized on open-source models transfer to black-box and proprietary models. 

\paragraph{Jailbreaks against vision language models.} \citet{carlini2024aligned} demonstrated that incorporating image inputs can enable more effective and efficient jailbreaking attacks. Concurrent to our work, \cite{schaeffer2024universal} explored whether optimizing jailbreak images on adapter-based vision language models exhibit the same transferability properties as GCG prompts. Their results indicate that jailbreak images are effective against adapter-based vision language models under white-box access but they do not generalize across models.

\section*{Impact Statement}
Our research contributes to the safety and responsible development of future AI systems by exposing limitations in current models. While acknowledging the potential for misuse in adversarial research, we believe our methods do not introduce any new risks or unlock dangerous capabilities beyond those already accessible through existing attacks or open-source models without safety measures. Finally, we believe that identifying vulnerabilities is essential for addressing them. By conducting controlled research to uncover these issues now, we proactively mitigate risks that could otherwise emerge during real-world deployments.

\bibliography{iclr2025_conference}
\bibliographystyle{iclr2025_conference}

\newpage

\appendix
\section{Effects of Softmax Temperature in Attack Success}
\label{ap:softmax}

Our 1-hot shortcut learns a very good mapping from VQ-VAE latent representations to the model input vocabulary. This implies that most predictions are very confident (i.e. probability 1 on the correct token and 0 elsewhere). This skewed distribution vanishes gradients for all other tokens and makes optimization harder. We found that increasing the temperature of the softmax helps distribute the probability over other likely tokens and make optimization smoother. Table ~\ref{tab:softmax} reports the results obtained with different choices of softmax. We report the numbers with and without the shortcut on the forward pass. We select a softmax temperature of 6 as default for all our experiments.

\begin{table}[ht]
    \centering
    \caption{Attack success rate with our 1-hot shortcut while using different temperature on the prediction softmax. Increasing the temperature improves effectiveness and transferability of the attack.}
    \label{tab:softmax}
    \begin{tabular}{@{} lrr @{}}
    \toprule
     Temperature    &  With Shortcut & Without Shortcut \\ \midrule
      1   &  12.9\% & 0.0\% \\
      2   &  25.6\% & 24.4\% \\
      4   &  34.7\% & 36.0\% \\
      6 (default)   &  \textbf{71.3\%} & 47.5\% \\
      8   &  46.8\% & 7.8\% \\
      10   &  25.6\% & 0.0\% \\ \bottomrule      
    \end{tabular}
\end{table}

\section{Transferability of Direct Attacks}
\label{ap:direct-transfer}

\begin{table}[ht]
\centering
\caption{\textbf{Direct attacks do not transfer across models.} We evaluate whether images optimized for a single prompt can jailbreak that prompt in different architectures.}
\label{tab:transfer-direct}
\begin{tabular}{@{} lrr @{}}
\toprule
Attack                                                             &  Chameleon-30B & LLaVA-1.6 \\ \midrule[\heavyrulewidth]
                                                        
No attack                                                                                                                                                  & 0.0\%         & 8.8\% \\ \midrule
Jailbreak Image (1-hot) & 0.0\% & 10.0\% \\
Jailbreak Image (Embed) & 0.0\% & 10.0\% \\
\bottomrule               
\end{tabular}
\end{table}

\end{document}